\documentclass[12pt]{article}
\begin{document}
\begin{flushright}
\end{flushright}
\begin{center}
{\large \bf Obtaining Bounds on the Sum of Divergent Series in Physics }
\end{center}

\vspace{0.5in}

\begin{center}
{Rajesh R. Parwani\footnote{parwani@nus.edu.sg}} 
\vspace{0.5in}

{University Scholars Programme,\\}
{Old Administration Block,\\}
{National University of Singapore,\\}
{10 Kent Ridge Crescent,\\}
{Singapore 119260.}

\vspace{0.25in}
\end{center}

\begin{abstract}
 
Under certain circumstances, some of which are made explicit here, one can deduce bounds 
on the full sum of a perturbation series of a physical quantity by using a variational Borel map on the partial series. The method is illustrated by applying it to various examples, physical and mathematical. 
\end{abstract}

\section{Introduction}

As there are hardly any problems in physically relevant quantum field theories that are exactly solvable, one must resort to approximate calculations. The only 
analytical tool that in principle allows for such approximate computations to be systematically improved, and  for the errors to be estimated, is perturbation theory. Unfortunately
in many cases the perturbation parameter is not small so that the
series, of which in practice only a few terms are known, diverges or gives a poor
representation of the physical quantity. Consequently, several techniques have
been used to estimate 
the full sum of an infinite series from the small number of given terms, see for example \cite{order,ellis,jen} and references therein. 

A variational method which was introduced in Ref.\cite{RP} has the novelty of apparently giving a monotonically converging sequence of approximations and thus possible bounds on the sum of an infinite series. That method has been applied to the physically important case of the free energy of gauge theories at high temperature in \cite{RP,rp3} and 
the purpose of this paper is to provide a detailed technical justification for the methodology used there. In essence this paper is a condensed and slightly modified version of the unpublished report \cite{unpub}.  The exposition here is not meant to satisfy a mathematical physicist but is rather suggestive and is illustrated with several examples to make some of the technical assumptions  plausible. It is hoped that some readers will find the method sufficiently interesting to re-examine it more rigorously in the future.  

In the following section, I describe in a
detailed but abstract manner how the technique of \cite{RP} can be used to obtain lower or upper
bounds for some series, and also in many cases to obtain accurate estimates of the
exact sum.
Concrete examples are given in Sec.(3). A conclusion and summary of open problems appears in Sect.(4). Finally, the Appendix contains 
an approximate but important analysis of the relevant equations which 
explains some of the empirically observed trends.

\section{The Method}

Consider the expansion
\begin{equation}
\hat{S}_{N}(\lambda)= \sum_{n=0}^{N} f_n \lambda^{n} \, ,
\label{ser}
\end{equation}
with $\lambda$ the perturbation parameter. It will be assumed in this paper that the relevant expansions are infinitely long, but that only a finite number of terms are known. Since perturbation expansions in quantum field theories have the 
generic behaviour $f_n \sim n!$ for $n$ large \cite{order}, it is 
natural to consider the Borel transform
\begin{equation}
B_{N}(z) = \sum_{n=0}^{N} {f_n \over n!} z^n \, .
\label{borel}
\end{equation}
The series (\ref{ser}) can then be recovered from the Borel integral,
\begin{equation}
\hat{S}_{N}(\lambda) = {1 \over \lambda} \int_{0}^{\infty} \
e^{-z/\lambda} \ B_{N}(z) \ dz \, .
\label{laplace}
\end{equation}
If the exact function $B(z) \equiv B_{\infty}(z)$ is known, 
Eq.(\ref{laplace})
can be taken
as defining the exact sum, $S=\hat{S}_{\infty}$, 
of the full perturbation series if the Borel integral
is well-defined.
The poor convergence of (\ref{ser}) can be attributed to the expansion of $B(z)$
in a power series as in (\ref{borel}) and the subsequent use of that series
beyond its radius
of convergence in (\ref{laplace}).

Suppose initially that $B(z)$ has only one singularity in the complex $z$-plane (the Borel
plane), at
$z=-1/p$, with $p$ real and positive. Then the radius of convergence of the Borel
series (\ref{borel}) is $1/p$. Therefore in order to construct an approximation to the
exact
sum, one must extend the domain of convergence of the partial series in
Eq.(\ref{laplace}).
The method of Loeffel, Le-Guillou and Zinn-Justin \cite{L,LZ} is to use a conformal map

\begin{equation}
w(z) = { \sqrt{1 + zp} -1 \over \sqrt{1 + zp} +1 } \, ,
\label{con}
\end{equation}
which maps the $z$-plane into a unit circle in the $w$-plane, with the
singularity
at $z=-1/p$  mapped to $w=-1$. The inverse of this map is

\begin{equation}
z(w) = {4w \over p} {1 \over (1-w)^2} \, .
\label{icon}
\end{equation}

Making the change of variables (\ref{icon}) in (\ref{laplace}) one obtains,

\begin{equation}
\hat{S}_{N}(\lambda) = {1 \over \lambda} \int_{0}^{1} \ dw \ {dz \over dw} \
e^{-z(w)/\lambda} \ B_{N}(z(w)) \, ,
\label{change}
\end{equation}
where  $B_{N}(z(w))$ is understood as a power series in $w$ obtained by an
expansion of
(\ref{icon}). In terms of the variable $w$, the series $B_{N}(z(w))$ converges
for
$|w|<1$, so that the potential problem in (\ref{change}) is only at the upper
limit of integration. Now, the difference between  $B_{N}(z(w))$ and
$B_{N+1}(z(w))$
begins at order $w^{N+1}$. If one knows the coefficients $f_n$ only 
up to $n=N$,
then it is consistent to keep only terms up to order $w^N$ in $B_{N}(z(w))$.
Performing this truncation in (\ref{change}) and riverting back to the $z$
variable,
one obtains

\begin{equation}
S_{N}(\lambda) \equiv {1 \over \lambda} \sum_{n=0}^{n=N} \ {f_n \over n!} \
\left({4 \over p}\right)^n \ \sum_{k=0}^{N-n}\
{(2n+k-1)! \over k! (2n-1)!} \ \int_{0}^{\infty} \ dz \ e^{-z/\lambda}\
 w(z)^{(k+n)} \, .
\label{resum}
\end{equation}
$S_N(\lambda)$ is a non-trivial resummation of the original series
$\hat{S}_{N}(\lambda)$.
In Ref.\cite{LZ,GZ}, Eq.(\ref{resum}) has been used to resum long perturbation
expansions for critical exponents, with $z=-1/p$ the location of the instanton
singularity of $\phi^{4}_{3}$ field theory. 
Note that after a scaling, (\ref{resum}) may be written as
\begin{equation}
S_{N}(\lambda) \equiv  \sum_{n=0}^{n=N}\ {f_n \over n!}\ \left({4 \over p}\right)^n \
\sum_{k=0}^{N-n}\
{(2n+k-1)! \over k! (2n-1)!} \ \int_{0}^{\infty}\ dz\
e^{-z} \ w(\lambda z)^{(k+n)} \, .
\label{resum2}
\end{equation}
Since $w(z)$ is a bounded and slowly varying function, this shows that the
resummed
expression $S_{N}(\lambda)$ is a much slower varying function of
the coupling than the original series $\hat{S}_{N}(\lambda)$.

The resummation (\ref{resum}) has also been applied to
quantum chromodynamics (QCD), with $z=-1/p$ the location of the first
ultraviolet renormalon pole \cite{qcd,qcd2}. 
Actually, QCD is Borel-nonsummable \cite{ben},
which means
that $B(z)$ also has poles for $z>0$, rendering the Borel integral 
ambiguous. In such a case, one can still use (\ref{laplace}) to define
the
sum of the perturbation series once the integral is made definite through some
prescription
such as the principal value.

In all the applications of (\ref{resum}) in \cite{LZ,GZ,qcd,qcd2}, $p$
is a known and fixed constant which determines the location of the singularity
of $B(z)$ closest to the origin. In Ref.\cite{RP}, the expression (\ref{resum})
was used as the starting point for a new technique which can be used even in 
cases when the singularity structure of $B(z)$ is unknown. Indeed, as seen above,
the conventional transformation of a series

\begin{equation}
\hat{S}_{N}(\lambda)= \sum_{n=0}^{N} f_n \lambda^{n} \, .
\label{sser}
\end{equation}
by the Borel-conformal map method results in the reorganised expression
\begin{equation}
S_{N}(\lambda,p) \equiv {1 \over \lambda}  \sum_{n=0}^{n=N} \ {f_n \over n!} \
\left({4 \over p} \right)^n \ \sum_{k=0}^{N-n} \
{(2n+k-1)! \over k! (2n-1)!} \ \int_{0}^{\infty}\  dz\  e^{-z/\lambda}\
 w(z)^{(k+n)} \, ,
\label{rresum}
\end{equation}
with $p$ a {\it fixed} constant. However, notice that $p$ does not figure in
(\ref{sser}) but enters (\ref{rresum}) only through the conformal map
(\ref{con}).
Thus instead of treating $p$ as some fixed constant as in Ref.\cite{LZ,GZ,qcd,qcd2},
one may consider $p>0$ a free parameter which defines the conformal map
(\ref{con}) and Eq.(\ref{rresum}) then represents a continuous family of
resummations,
one for each value of $p$. {\it Thus, from now on, $p$ will not refer
to the location of some singularity in $B(z)$}. In fact, except for the
regularity of $B(z)$ at the origin, no other knowledge about
the singularity structure of $B(z)$ is required 
for the  method elaborated below.

Of course having liberated ourselves from the usual interpretation
of $p$ in (\ref{rresum}), we also lose definiteness in our resummation.
Therefore
some new condition must be imposed to fix the value of $p$ and hence 
of the expression (\ref{rresum}). For a start, 
for each $N$, choose
$p>0$ to be the location of an extremum of $S_{N}(\lambda,p)$ \cite{RP}.
Since (\ref{rresum}) depends on $\lambda$, the value of $p$ in principle
will also depend on $\lambda$. However the procedure then becomes too
unwieldy, and to simplify it $p$ is determined at
some reference
value $\lambda=\lambda_0$, say at the mid-point of the range of interest:

\begin{equation}
{\partial S_{N}(\lambda_o, p) \over \partial p} =
0 \, .
\label{ext}
\end{equation}
As mentioned after
(\ref{resum2}), $S(\lambda,p)$ is a relatively slowly varying function of
$\lambda$,
hence determining $p$ at some fixed $\lambda=\lambda_0 $, and then using the
same $p$
in (\ref{rresum}) for various $\lambda$ is sufficient for practical purposes.
Indeed,
in many applications, one requires the sum of the series at one particular
value of the coupling, or in a very narrow range, so the simplification made in
(\ref{ext}), is both useful and sufficient.

In some cases, (\ref{ext}) will not have any solutions. For example, if
all the $f_n$ are of the same sign, $S_{N}(\lambda,p)$ will be a monotonic
function
of $p$. Thus for the procedure to work, at least some of the $f_n$ must be of
a different sign. { \it This will be assumed to be the case from now on}.
Suppose next that
$ f_{1} \neq 0$. Then for $p$
large,
and since the $p$ dependence of $w(z)$ is mild,

\begin{equation}
S_{N}(p) \sim f_0 + {f_1 \over p} \, .
\end{equation}
Therefore for $f_{1} <0$, $S_N$ will first decrease as $1/p$ increases and
then increase when the next term $f_m/p^m >0$ dominates the sum. 
If the last non-zero term $f_N / p^N$ is positive then one expects $S_N(p)$ 
to have a global minimun at some $p>0$. However if $f_N / p^N <0$ then $S_N(p)$
can become very small as $p \to 0^{+}$ and so the minimum would likely be only a local extremum.
From this heuristic argument one concludes that if the first non-zero coefficient $f_n, (n>0)$
is negative, Eq.(\ref{ext}) will have global minima solutions for some $N$.   

For conciseness,
unless otherwise stated, I will discuss from now on only the case when a global
minimum exists,
as the arguments for the global maximum case are then obvious.
{ \it For a series $S_N(\lambda,p)$, define $p(N)$ to be the position of the global minima for 
the values of
$N$ when they exist, and for other values of $N$ let $p(N)$ denote the 
location of the local minima}. Also, let

\begin{equation}
S_{N} \equiv S_{N}(\lambda, p(N)).
\label{sn}
\end{equation}

Obviously the definition of $p(N)$ has been chosen because it is useful. If $p(N)$ is a location of 
a global minimum, then $S_N$ certainly is a lower bound on $S_N(\lambda,p)$. However this does not imply that
$S_N$ is a lower bound on the exact sum $S$ of the pertutbation series  because for finite $N$, 
$S$ might lie outside of the space of resummations labelled by $p$.
For each $N$, let $p^{\star}(N)$ be the value of $p$ that is optimal,
that is, it is the value which when used in (\ref{rresum}) gives the best
estimate of $S$.
Define, $S^{\star}_{N} = S_{N}(\lambda,p^{\star}(N))$. Then for a global minima one
has 
\begin{equation}
S_{N} \leq S_{N}^{\star}
\label{star}
\end{equation}
Presumably $S_{N}^{\star}$ converges to $S$ as $N \to \infty$. Then
for those $N$ when global minima exist,
\begin{equation}
S_{N \to \infty} \leq S \, .
\label{limit}
\end{equation}
(This implicitly assumes that the sub-sequence of global minima is infinite: That is, 
given any positive integer $N_0$, there is some $n >N_0$ for which $f_n$
 is positive.)

Though the inclusion of global minima in the definition of $p(N)$ is quite clear, the
inclusion of local minima requires some explanation. As mentioned earlier, if $S_N(\lambda.p)$
has a global minimum, then $S_{N+1}(\lambda,p)$ will probably not have a global minimum if $f_{N+1}$
is negative because $S_{N+1}(\lambda,p)$ might become very small as $p \to 0^{+}$. 
However $S_{N+1}(\lambda,p)$ will still have a local minimum (and hence also a local maximum).
The local minimum will occur for moderate values of $p$ close to $p(N)$, the global minimum position of 
$S_N(\lambda,p)$. Therefore one might expect that the local minimum for $S_{N+1}(\lambda,p)$ still gives a bound on 
$S^{\star}_{N+1}$. Indeed, suppose that the inequality
\begin{equation}
S_{N} \leq S_{N+1}
\label{ineq}
\end{equation}
holds for {\it all} $N$ up to $N=\infty$, then clearly
\begin{equation}
S_{N} \leq S
\end{equation}
and one concludes that $S_N(\lambda, p(N))$ is a lower
bound on the sum of the full perturbation series. Of course proving
(\ref{ineq}) is equivalent to knowing $f_n$ explicitly for all $n$, which is not the
situation in reality. As a practical matter, it is sufficient to draw a
conclusion
after observing the trend (\ref{ineq}) (or lack of) for the available terms of
the series. Arguments given in the Appendix indicate that the trend, once started, will
continue. 

An interesting question is whether the inequality (\ref{limit}) is saturated. In most of
the examples studied this has been found to be the case. 
In fact the convergence
is so rapid that one can conclude with a high degree of confidence, not only
that $S_N$ is a lower bound, but that it is close to the exact value. However 
the example in Sec.(3.5) shows that the series $S_N$, 
though forming a bound on the actual value $S$, might not converge to $S$. 
In that example the exact result $S(\lambda)$ has a first derivative 
$d S(\lambda) /d \lambda$ that varies rapidly with $\lambda$. 
An explanation of why in such cases  
the bounds $S_N$ do not converge to the exact value is given in the
Appendix. However, as described below, 
such situations can also be taken care of 
by the resummation procedure (\ref{rresum}-\ref{ext}). 

Recall that
when a local minima occurs for some $N$, one expects also a local maximum. 
{ \it Define $\bar{p}_{N}$ to be the position of those local maxima and $\bar{S}_N$ the 
corresponding value of the resummed series}. It turns out that $\bar{S}_N$ also
satisfies an inequality like (\ref{ineq}) (see Sec.(3.5)).
However since for the sequence $\bar{S}_N$ of only {\it local} extrema 
one does not seem to have a statement like 
(\ref{star}-\ref{limit}), it is not {\it a priori} obvious that they form bounds
to the exact result. Of course if one believes that the $p^{\star}(N)$
as defined above always take moderate values, then a statement like (\ref{star}) might also
be made for the sequence $\bar{S}_N$. 

Furthermore, given a series which has global minima solutions to (\ref{ext}), then by 'removing' the first nontrivial
coefficient of that series one forms an auxiliary series, $S^{'}_{N}(\lambda,p)$, which has 
global maxima solutions. In this way one can 
form alternative bounds to the sum of the original series 
complementing those obtained from $S_N$. This is illustrated in Sec.(3.5). 

Using the sequences $S_N$, $\bar{S}_N$,
and $S^{'}_{N}$, one can obtain constraints on $S$ and, with some physical input, an estimate of
$S$ itself. A question now arises for practical problems where exact results are { \it not known}. Is it possible to determine whether the unknown exact result lies closer to the
upper or lower bound ?  Empirically, it appears that the series $S_N$ converges to the 
exact result $S(\lambda)$ if $\partial^2 S(\lambda) / \partial \lambda^2$ is very small
in magnitude in the entire range of interest, $0< \lambda < \lambda_0$. Otherwise the
exact result seems to be better approximated by the series $\bar{S}_N$ or by 
the auxiliary series method discussed above. An explanation of why the bounds
$S_N$ have slowly varying first derivatives, and why the bounds $\bar{S}_N$
and those from the auxiliary series have faster varying first derivatives
is given in the Appendix.   

Although the main discussion in this paper will be for the $p(N)$ (or $\bar{p}(N)$ and $p'(N)$) 
as defined above, in some cases one finds (by inspection) solutions $p_0(N)$ to (\ref{ext})
which have the property that as $N \to \infty$,\ $p_0(N) \to p_0$, a constant.
In such a case one is tempted to speculate that the fixed point $p_0$
indicates the existence of a singularity at $z=-1/p$ in $B(z)$. This is indeed
found to be the case in the examples studied, though apparently the converse is
not necessarily true: the singularities of $B(z)$ need not show up as solutions of 
(\ref{ext}). Furthermore, the convergence of the series $S_{(0)N}$ is not expected to be
monotonic since in general the $p_0(N)$ refer to both maxima and minima.

Though not manifest at first sight, the analysis in the Appendix suggests that even 
the $p(N)$ defined above Eq.(\ref{sn}) actually will converge to a fixed value
as $N \to \infty$. This trend can be observed in the examples studied.

\section{Examples}

In this section a number of examples with different properties are studied
to illustrate the resummation technique of Eqs.(\ref{rresum}-\ref{ext}) and to show that the various technical assumptions made there are not vacuous.
Except for the last example, all define Borel-summable series,
that is, $B(z)$ has no singularities at real, positive $z$. A Borel-nonsummable
example will be considered in Sec.(3.3). All of the examples here have global
minima as solutions to the extremum condition (\ref{ext}) for some $N$. (From any such
series $S_N$ one can trivially construct $C-S_N$, with $C$ a constant,
which then gives an example of a series with global maxima solutions 
to (\ref{ext})). 
The reader is reminded
that the existence of global minima  does not by itself imply
that one has obtained a lower bound on the sum of the series.
The additional
condition (\ref{ineq}) must be satisfied for a lower bound. The Euler-Heisenberg series below shows how 
one can end up with an upper bound from global minima ! (See also the Appendix).

It is also re-emphasized that although in these  examples the exact singularity
structure of $B(z)$ is known, that information will {\it not} be used in the
resummation. The resummation of the partial series 

\begin{equation}
\hat{S}_{N}(\lambda)= \sum_{n=0}^{N} f_n \lambda^{n} \,
\end{equation}
will proceed using Eqs.(\ref{rresum}-\ref{ext}).
The exact $B(z)$ will only be used to compare the resummed results with the
exact sum of the full series given by the Borel integral

\begin{equation}
S(\lambda) = {1 \over \lambda} \int_{0}^{\infty} e^{-z/\lambda} B(z) dz \, .
\label{integral}
\end{equation}

In addition to the examples below, the interested reader may find others in \cite{unpub}.

\subsection{A Prototype}

Consider a model defined by the Borel function,

\begin{equation}
B(z) = { 1 \over 1+z}   \, .
\label{eb1}
\end{equation}
Expanding $B(z)$ to $N$-th order in $z$ and using the result in (\ref{laplace})
gives the truncated series

\begin{equation}
\hat{S}_{N} = \sum_{n=0}^{N} (-\lambda)^n \ n! \; .
\label{es1}
\end{equation}
The divergent nature of this series is displayed in Fig.(1a).
Using $f_n = (-1)^n  n!$ in (\ref{rresum}) and solving Eq.(\ref{ext}) at the
reference
value $\lambda_0=1$ gives the following solutions (minima): $p(2)=2.65,\
p(3)=5.1$ and $p(4)=8.4$. As expected from the arguments of Sec.(2),
there is no solution for $N=1$, the solutions for $N=2$ and $N=4$ are global minima
while that for $N=3$ is a local minimum. The resummed series is shown in
Fig.(1b).
The convergence of the $S_N$ is monotonic and satisfies the condition (\ref{ineq}). 
Hence the approximants
$S_N$ can be argued to form lower bounds to the exact result. That this is
indeed
the case can be seen by inspection of the exact result in Fig.(1b) as obtained
from (\ref{eb1}) and (\ref{integral}). Furthermore, it is clear that the lower
bounds
converge to the exact value, and so may be used to estimate it. 
At $\lambda=0.5$, the exact value is $0.722657$, while the approximants
are $S_2=0.704, \ S_3=0.709, \ S_4=0.711$.

For this model there are other solutions, for each $N$,
to the extremum condition
(\ref{ext}) in addition to the minima. 
By inspection one picks out the sequence of values, $p_0(3)=1.6$ (local maximum), $p_0(4)=1.3$
(local minimum) and $p_0(5)=1.15$ (local maximum) as plausibly approaching a fixed point. 
Indeed we already know
that in this model the exact singularity of $B(z)$ is a pole at $z=-1/p=-1$,
 so the
fixed point probably refers to the location of this singularity. It should come as
 no surprise that
if the approximants (\ref{rresum}) are evaluated at the values $p_{0}(N)$,
 the convergence to the
exact value will be much faster, and this is indicated in Fig.(1c). At 
$\lambda=0.5$ the values of the approximants are $S_{(0)3}=0.726,\ S_{(0)4}=0.7219, \ S_{(0)5}=0.7228$.
Notice however that in this case the convergence is not monotonic (and was not expected to be).

\subsection{Beta function of a Two-Dimensional Field Theory}

In \cite{JJ} the exact beta function in a two-dimensional field theory was obtained and the result checked explicitly up to three-loop order. After a scaling, that beta function is essentially of the form 

\begin{equation}
S(\lambda)={1 \over 1+ \lambda} \ .
\label{eb31}
\end{equation}
In this case a power expansion
of $S(\lambda)$ is actually convergent for $|\lambda| <1$.
However close to $\lambda=1$,
the convergence is very slow and one
requires a large number of terms of the series to obtain accurate results.
The truncated perturbation series corresponding to (\ref{eb31}) is

\begin{equation}
\hat{S}_{N} = \sum_{0}^{N} (-1)^n \ \lambda^n  \, .
\label{es3}
\end{equation}
Notice that there is no factorial growth of the coefficients in (\ref{es3}).
The slow convergence of this series for $\lambda$ close to $1$
is displayed in Fig.(2a).

The utility of the resummation procedure in this case is now demonstrated.
Using $f_n = (-1)^n$ in (\ref{rresum}) and solving Eq.(\ref{ext})
 at the reference
value $\lambda_0=1$, gives the following solutions (minima):
$p(2)=1.2,\
p(3)=2.9, \ p(4)= 5$. The resummed series is shown in Fig.(2b)
together with the exact result.
The convergence is rapid, monotonic, satisfies the condition (\ref{ineq}),
and the approximants $S_N$ form lower bounds to the exact result.
With only four terms, the resummed series already shows a dramatically
improved
convergence compared to the original series (\ref{es3}), even for
couplings as large as $\lambda =0.8$: The exact sum at that coupling is
$0.556$, while the resummed values are $S_2=0.492, \ S_3=0.507,\ S_4=0.512$.

This example illustrates that the resummation
procedure (\ref{rresum}-\ref{ext}) is useful
even for a convergent series, especially when  
one is close to the
radius of convergence and when only a few terms of the series are 
available (which is often the situation in practice). 
Furthermore this example emphasizes that the parameter $p$ has no obvious relation to the singularities of $B(z)$. Indeed the function (\ref{eb31}) corresponds to the Borel transform 
\begin{equation}
B(z) =  e^{-z}  \, .
\label{eb3}
\end{equation}
which is regular everywhere in the Borel plane.

\subsection{A Borel-Nonsummable Model}

For most of the physical quantities calculable from the Standard Model of
particle physics, the perturbation expansion is not expected
to be Borel summable. In simple terms, this means that the function
$B(z)$ has poles on the positive semi-axis of the Borel plane thus
rendering the Borel integral (\ref{integral}) ambiguous.
If one choses an $i \epsilon$ presciption then the resulting
ambiguity is in the imaginary part. Sometimes these
imaginary parts are of direct physical relevance \cite{dunne,jen}. 
More generally
they indicate that the perturbation series does not give the full answer,
but must be supplemented with some non-perturbative contributions 
\cite{qcd,ben}.
Since the imaginary parts will be of the form $e^{-1/\lambda}$,
the additional {\it real} non-perturbative terms are expected
to take the same form. These expectations have been confirmed in some
lower dimensional field theories (see \cite{ben} and references therein).

One can also construct mathematical models to illustrate the arguments of the
last paragraph. Suppose, for simplicity, that the only
singularity of $B(z)$ for $z>0$ is a single pole at $z=q$. Then the sum of the
perturbation series can be defined by the principal value prescription,

\begin{equation}
S_{pert} \equiv {1 \over \lambda} {\cal P}\int_{0}^{\infty} \ dz \
e^{-z /\lambda}\ B(z) \, .
\label{prin}
\end{equation}
With this definition, one focuses only on the real part of the physical
quantity. Suppose furthermore, again for simplicity,
that $B(z)$ is integrable at infinity. Then the Eq.(\ref{prin}) may be rewritten
as

\begin{equation}
S_{pert} = {1 \over \lambda} \int_{0}^{\infty} \ dz \
 (e^{-z/\lambda}-e^{-q/\lambda}) \ B(z) \; + \;
{ e^{-q/\lambda} \over \lambda} {\cal P} \int_{0}^{\infty} \ dz \ B(z) \, .
\label{rewrite}
\end{equation}
The first term on the right-hand-side is finite and unambiguous. Call it
$S_{exact}$, and denote the second term on the right-hand-side as $S_{np}$.
Thus we have

\begin{equation}
S_{exact}(\lambda) = S_{pert}(\lambda) - S_{np}(\lambda) \, .
\label{reorg}
\end{equation}

In this way, the exact result $S_{exact}$,
has been broken into two components, $S_{pert}$ which is purely perturbative,
and
$S_{np}$ which is purely non-perturbative. However while $S_{exact}$ is
well-defined, both $S_{pert}$ and $S_{np}$ are only defined through the
principal value prescription. Though highly simplified, this model plausibly
represents the situation, for example, in Quantum Chromodynamics (QCD).

In order to illustrate explicitly
the resummation technique (\ref{rresum}-\ref{ext})
in the Borel-nonsummable case, set
\begin{equation}
B(z) = {1 \over (1+z)(5-z)} \, .
\label{eb7}
\end{equation}
Then,
\begin{equation}
S_{pert}(\lambda)= {1\over \lambda} {\cal P} \int_{0}^{\infty} {e^{-z/\lambda}
\over (1+z)(5-z)} \,.
\label{p6}
\end{equation}
The truncated perturbation series corresponding to (\ref{p6}) is
\begin{equation}
\hat{S}_{N} = \sum_{0}^{N} f_n \lambda^n   \, ,
\label{es7}
\end{equation}
with
\begin{equation}
f_n= ((-1)^n   +   5^{-(n+1)}) \ {n! \over 6} \, .
\end{equation}
The divergent series is oscillatory, analaogous to that Fig.(1a).
The solution of Eq.(\ref{ext})
 at the reference
value $\lambda_0=1$, gives (minima):
 $p(2)=2.8,\
p(3)=5.4,\ p(4)=9, \ p(5)=13.5$.
The resummed partial series is shown in Fig.(3a) together with the exact
 sum of the full
{\it perturbation} series given by (\ref{p6}).
The convergence is rapid, monotonic, satisfies the condition (\ref{ineq}),
and the approximants $S_N$ do indeed form lower bounds. Also, the
 convergence is so fast that, except for a small interval at intermediate
 coupling, $S_5$
is not only a lower bound, but also a very good approximation to $S_{pert}$.

Now, corresponding to the Borel function (\ref{eb7}), one has from the
 definitions before Eq.(\ref{reorg}),
\begin{equation}
S_{np}(\lambda) =  {\ln{5} \over 6 \lambda} \ e^{-5/\lambda} \,
\end{equation}
and
\begin{equation}
S_{exact}(\lambda)= {1\over \lambda} \int_{0}^{\infty}
 {e^{-z/\lambda}-e^{-5/\lambda} \over (1+z)(5-z)} \,.
\end{equation}
In Fig.(3b), the curves for $S_{exact}$ and $S_{5}$ are plotted for 
a bigger range of $\lambda$. This shows
that except for a small range of couplings, the resummed perturbative
 approximant $S_5$ lies {\it above} and 
deviates significantly from the exact result
although, as seen above,
$S_5$ does form a converging lower bound to the { \it perturbative} component $S_{pert}$ of
the exact result. The difference is of course the non-perturbative piece
$S_{np}$. In the same figure there is a plot of $(S_{5} - S_{np})$ which,
according to the definition (\ref{reorg})
should  approximate $S_{exact}$.
Indeed the agreement is very good and, amusingly, it improves at large coupling: 
At $\lambda=5$, $S_{exact}=0.0533$ and $S_5 -S_{np} = 0.0457$, while at $\lambda=10$,
$S_{exact}=0.0219$ and $S_5 - S_{np}=0.0193$. 

This toy model discussion illustrates concretely the arguments given
in Ref.\cite{RP} for the free-energy density of thermal $SU(3)$ gauge theory.
There it was found that the exact result, given by lattice data,
 differed significantly from the resummed perturbative result,
 and it was argued that the difference was caused by the Borel-nonsummability
 of the theory. Assuming a non-perturbative component of the
 form ${A \over \lambda} \ e^{-q/\lambda}$, the constants $A$ and $q$ were
 determined from the difference between $S_{exact}$ and $S_5$. A more detailed
discussion of the results in \cite{RP} and their extension to full QCD and other
non-Abelian gauge theories is presented in \cite{rp3}.

\subsection{The Euler-Heisenberg Series}

Schwinger's \cite{sch} effective Lagrangian 
(density) for QED in a uniform magnetic field $B$ is,
\begin{equation}
L(\lambda) = -{e_{r}^2 B^2 \over 8 \pi^2} \int_{0}^{\infty} {ds \over s^2} \ \left( \mbox{Coth}(s)-{1 \over s} -{s \over 3} \right) \
e^{-s/\sqrt{\lambda}}
\label{sch}
\end{equation}
where $\lambda \equiv e_{r}^2 B^2/m^4$, $e_{r}$ is the renormalized electron coupling, and $m$ the electron mass.

Define the dimensionless quantity 
\begin{equation}
S(\lambda) = 100 \int_{0}^{\infty} {ds \over s^2} \ \left( \mbox{coth}(s)-{1 \over s} -{s \over3}\right) \
e^{-s/\sqrt{\lambda}}
\label{sch2}
\end{equation}
which is related in an obvious way to (\ref{sch}). An expansion of (\ref{sch2}) is given 
by \cite{dunne}

\begin{equation}
\hat{S}(\lambda) = 1600 \sum_{n=1}^{\infty} f_n \lambda^n \,
\label{EH}
\end{equation}
with
\begin{equation}
f_n = { 4^{n-1} \ {\cal{B}}_{2n +2}  \over 2n \ (2n+1) \ (2n+2)} \, ,
\end{equation}
and where ${\cal{B}}_{2n}$ are the Bernoulli numbers which alternate in sign, and diverge factorially for large $n$. 
Equation ({\ref{EH}) is the Euler-Heisenberg \cite{eh} expansion which is equivalent to a sum of an infinite number of
Feynman diagrams consisting of one closed fermion loop with an even number of external photons. 
The Euler-Heisenberg series diverges for large values of 
$\lambda$ and displays oscillatory behaviour analogous to that in Fig.(1a). Consider now a resummation of the divergent series using (\ref{rresum}-\ref{ext}). 
At the reference value $\lambda=10$, the solutions to (\ref{ext}) are (global minima) $p(2)=0.77, \ 
p(3)=1.4,\  p(6)=2.25$.  No solutions were found for $N=3,5,7$.

The exact result of Schwinger, given by (\ref{sch2}), is plotted in Fig.(4), together with the approximants
$S_2, \ S_4, \ S_6$.  The approximants form {\it upper} 
bounds although the $p(N)$ are positions of global {\it minima}. See again the caution prior to
Eq.(\ref{star}) and the Appendix. The convergence of the approximants $S_N$ to the exact expression
is manifest. For example, at $\lambda=10$, where the Euler-Heisenberg series is badly divergent,
the Schwinger's exact value is $S(10)=-8.056$ while the estimates are, $S_2=-5.9, \ S_4=-6.9, \ S_6=-7.3$.

The Euler-Heisenberg series (\ref{EH}) is Borel summable \cite{dunne}, and
therefore provides an example of a Borel-summable series in a theory (QED) which 
is generally considered Borel-nonsummable. However it should be noted that 
the Euler-Heisenberg series (\ref{EH}) represents a one-fermion-loop result
whereas the "renormalon" singularities \cite{ben} which signal Borel-nonsummability 
are expected to arise when  multi-fermion loop corrections to 
(\ref{sch}) are taken into account.

\subsection{Critical Exponents in Three Dimensions}

The three-dimensional $O(\cal{N})$ symmetric $\phi^4$ field theory can be used as an effective theory 
(see \cite{LZ,GZ} and references therein) to describe the critical behaviour of many physical systems 
near a second-order phase transition. For this purpose, the renormalization group functions of this theory
have been calculated to very high order. A detailed list of references can be found in \cite{GZ,AS}.   
For example, the beta function for the polymer case, ${\cal{N}}=0$, is given by 
\begin{equation}
\beta(\lambda)= -\lambda + \lambda^2 - 0.439815 \lambda^3 + 0.389923 \lambda^4 - 0.447316 \lambda^5
+0.633855 \lambda^6 - 1.03493 \lambda^7 \, ,
\label{beta}
\end{equation}
where $\lambda$ is a dimensionless coupling (usually denoted as $g$ or $\tilde{g}$
 in the literature \cite{GZ}). The objective is to 
find the non-trivial infrared fixed point, $\lambda^{\star}$, of the
theory, which is given by the zero of the beta function,
\begin{eqnarray} 
\beta(\lambda^{\star}) & =& 0 \, , \\
{\partial \beta \over \partial \lambda}|_{\lambda^{\star}} & \equiv &  \omega >0 \, .
\end{eqnarray}
The expression (\ref{beta}) is divergent for large $\lambda \sim 1$ 
where a nontrivial zero is expected. Using the resummation (\ref{rresum})
with $S_N$ denoting approximants to the beta function, and choosing the reference point 
$\lambda_0=1$, one gets as solutions to 
Eq.(\ref{ext}) (minima), $p(2)=1.3, \ p(3)=3.2, \ p(4)= 5.6, \ p(5)=8.6, \ p(6)=12.25, \
p(7)=16.5$. The curves are shown in Fig.(5a). They form rapidly converging 
lower bounds but intercept the $\lambda$-axis only at the origin, thus giving
only a trivial zero to the beta function. The situation is similar to the
$\mbox{sin}(\pi \lambda)$ toy model studied earlier.

Now, since for $N=3,5,7$ the minima are only local, one expects local 
maxima to exist also.
Indeed the local maxima are located at, $\bar{p}(3)=0.18, \ \bar{p}(5) = 0.21, \ 
\bar{p}(7)=0.19$. The curves for $\bar{S}_N$ are shown in Fig(5b). Though these are
local maxima, they obey the inequality $\bar{S}_{N+1} > \bar{S}_{N}$ and so 
appear to form lower bounds ! Since the lower bounds due to 
$\bar{S}_N$ are {\it higher} than the lower bounds due to $S_{N}$ (Fig.(5a)), 
one may argue that
those due to $\bar{S}_N$ provide more accurate information. The curves in Fig.(5b) do in fact 
have a non-trivial zero. For $N=7$, the zero is near $\lambda=1.425$. It is reasonable to suppose that as {\it both} the approximants $S_N$ and $\bar{S}(N)$
provide lower bounds, it is simply a matter of choosing the highest lower bound to estimate the sum of the
series.

Given the physical importance of this example, it is useful to perform further checks. 
So let us re-analyse the problem using the auxiliary series method discussed in Sect.(2). 
That is,
let $S^{'}_{N}(\lambda,p) = S_{N}(\lambda,p) / \lambda$. The auxiliary series will then 
obviously have maxima as solutions to (\ref{ext}). The solutions at the reference point 
$\lambda_0=1$ are
$p'(2)=0.6, \ p'(3)=0.875, \ p'(4)=0.33, \ p'(5)=0.4, \ p'(6)=0.2525$. (Actually, 
$p'(5)$ is a point of inflexion). The curves for $\lambda S^{'}_{N}$ are shown in 
Fig.(5c). Again, though determined by positions of maxima, the curves appear to form
rapidly
converging {\it lower bounds} which have a non-trivial zero. 

In order to compare with the results in the literature, more precise numbers are now quoted.
Conjecturing that the curves in Fig.(5c) are indeed lower bounds to the exact result,
the $N=6$ curve, shown magnified near its nontrivial zero in Fig.(5d), gives 
an upper bound of $1.4193$ to the non-trivial zero of the beta function. Re-optimizing the
$N=6$ equation (\ref{ext}) at $\lambda_0=1.419$ does not change the curve for 
$\lambda S^{'}_{N}$ significantly to modify that bound at the level of accuracy quoted.
The slope of the beta function at the non-trivial zero can also
be determined from the $N=6$ curve in Fig.(5d). It is $\omega=0.7955$. Since the curves 
appear to get steeper as $N$ increases, this value of $\omega$ is a lower bound. 
These values can now be compared
with those of Ref.\cite{GZ}: there it was found that $\lambda^{\star}=1.413 \pm 0.006$ and $\omega=0.812 \pm
0.016$. The agreement with the bounds found here is excellent. (Comparison of
results obtained by other methods and other authors may be found in \cite{GZ}). 

It is important to note the difference between the methodology used in this paper
and that employed in Ref.\cite{GZ}. In \cite{GZ}
the authors used a Borel-Leroy transformation
with a variable parameter $b$ but they used the conventional conformal map \cite{LZ} 
with the value of $p$ {\it fixed} at
the precise location of the instanton singularity, $p^{\star}=0.166246$ 
(the value for ${\cal{N}}=0$). 
The parameter $b$, together with some other
parameters introduced in \cite{GZ} were used to check the convergence of the series
and to estimate their errors. Here instead the usual Borel tranform (\ref{borel}) is used
but the conformal map has a single variational parameter $p$ determined according to the
condition (\ref{ext}). From the numbers 
quoted above, it is clear 
that the values $p(N)$ used here are not the same as the value $p^{\star}$. 
Furthermore, 
in the approach of this paper,
the only assumptions made 
about the analyticity structure of the Borel transform 
are (i) $B(z)$ should be regular at the origin, and (ii)  
in order for
the resummed perturbation series to faithfully 
represent the physical quantity, the series 
should be Borel summable.
Borel summability of the $\phi^{4}_{3}$ theory 
has been established in \cite{ems}.

From this example it is clear that the novel resummation presented here,
with the parameter $p$ determined from (\ref{ext}), 
can be used to complement the analysis done in \cite{GZ}.

\section{Conclusion}

The method used in Ref.\cite{rp3} to resum the perturbative free energy of gauge theories at high temperature has been detailed. It can be used more generally to obtain  
bounds on the full pertubation expression, $S(\lambda)$, of a physical quantity  
if in addition to the given partial perturbation series
\begin{equation}
\hat{S}_N(\lambda) = \sum_{n=0}^{N} f_n \lambda^n \, ,
\label{partial}
\end{equation}
{\it some additional requirements are met and some assumptions made}.

The first requirement is that not all the $f_n$ be of the same sign. Then solutions
exist to the extremum condition (\ref{ext}). If the first nontrivial coefficient
$f_n (n>0) $ is negative, then the solutions to (\ref{ext}) will be global or local minima,
depending on the sign of $f_N$. The approximants $S_N$ as defined through (\ref{sn})
then form lower bounds to $S(\lambda)$ if the inequality (\ref{ineq}) 
is satisfied for all $N$ and {\it if the assumptions leading to (\ref{limit}) are admitted}. The inequality (\ref{ineq}) must first be explicitly tested for 
the available terms of the series, then the
arguments given in the Appendix suggest that the trend will
continue for those cases where $c(N)>1$ at low orders. 
Therefore even with partial information as in (\ref{partial}), 
one can apparently deduce
lower bounds to the exact value $S(\lambda)$ if the technical assumptions mentioned above are valid. Under similar conditions, additional bounds may be obtained by using the
auxiliary series method described in Sec.(2).

Sometimes one finds {\it minima} solutions to (\ref{ext}) but for which the $S_N$ obey an
inequality {\it opposite} to (\ref{ineq}). In such cases, even though the convergence is 
monotonic, it is not {\it a priori} obvious that
the $S_N$ are actually upper bounds to the exact value.  For problems where the exact result is
unknown, additional input from theory or physics is required before a definitive 
statement can be made. However in all the examples encountered of this type,
the $S_N$ did actually bound the exact result. A similar caveat concerns the 
approximants $\bar{S}$ defined near the end of Sec.(2).

The observed rapid convergence of the bounds has been explained in the Appendix.
The obvious question is whether the bounds converge to the exact value itself. 
Empirically, it is found that the bounds formed by $S_N$ actually converge 
to the exact value if $S(\lambda)$ has a slowly varying first derivative,
$\partial S(\lambda) / \partial \lambda$. Otherwise the complementary bounds formed by
$\bar{S}(\lambda)$ or the auxiliary series method give better approximations to the
exact result. The reason for this phenomena has been given in the
Appendix.

Although most of the observed features of the resummation method developed
here have been explained in the Appendix, at least in a semi-quantitative way,
more patterns were detected than could be explained. For example, (i) Is it true that in all cases, as $N \to \infty$, $c(N) \equiv p(N+1)/p(N) \to 1$ (Similarly for the 
$\bar{p}(N))$ ? 
(ii) Is it true that in all cases the approximants $S_N$ and $\bar{S}_N$ form
bounds to the full perturbative result ? (iii) Is it true that 
the bounds $S_N$ always approximate the
exact result $S$ well if $| {\partial^2 S \over  \partial \lambda^2} / {\partial S \over  \partial \lambda} |$ is small
throughout the range, $0< \lambda <\lambda_0$, of interest, and otherwise the
alternative bounds $\bar{S}_N$ or those from the auxiliary series give 
better approximations ?

Clearly it would be desirable to have a better understanding of the conditions  
under which some of the technical assumptions made above hold rigorously. Until that is achieved, and in particular for those physical cases where little is known about the exact series anyway, the method can nevertheless be used as another independent estimator of the sum of a perturbation series, complementing other techniques in the literature. I point out that the resummation method discussed here can be viewed as an example of an "order-dependent mapping" introduced in \cite{SZ} by Seznac and Zinn-Justin, but applied here to the Borel series rather than the original perturbation expansion. In addition, the extremum condition (\ref{ext}) can be thought of as an example of
Stevensen's Principal of Minimum Sensitivity \cite{S}.\\ \\

\vspace{0.3in}
\noindent
{\large \bf Acknowledgements}: I am grateful to C. Coriano$'$, S. Pola, J.S. Prakash,
A. Goldhaber, P. Van Nieuwenhuizen, I. Parwani and U. Parwani for hospitality during the course of this work.

\newpage

\noindent
{\Large {\bf Appendix}}\\
{\large {\bf  A1}}

The resummed perturbation series with a variable parameter $p$ is given by
\begin{equation}
S_{N}(\lambda,p) \equiv   \sum_{n=0}^{n=N} \ {f_n \over n!} \
\left({4 \over p} \right)^n \ \sum_{k=0}^{N-n} \
{(2n+k-1)! \over k! (2n-1)!} \ \int_{0}^{\infty}\  dz\  e^{-z}\
 w(z \lambda)^{(k+n)} \, .
\label{a01}
\end{equation}
Write the equation above in the compact form
\begin{equation}
S_N(\lambda,p) = \sum_{n=0}^{N} {A_n (N) \over p^n} \, ,
\label{a02}
\end{equation}
where $A_n(N)$ is a $p$ and $\lambda$ dependent constant that can be read off from (\ref{a01}). 
In particular note that the sign of $A_n(N)$ is the same as the sign of $f_n$.
The extremum condition (\ref{ext}) applied to (\ref{a02}) results in 
\begin{equation}
\lambda \ { \partial S_{N} \over \partial \lambda}|_{\lambda_0} = 
\sum_{n=1}^{N} {n A_n(N) \over p(N)^n} \, ,
\label{a1}
\end{equation}
where use has been made of the form $w(z \lambda)$ in transforming
the $p$ derivative of $A_n(N)$ into a $\lambda $ derivative. 
At the solution 
$p=p(N)$ of (\ref{ext}), one has  
\begin{equation}
S_N \equiv \sum_{n=0}^{N} {A_n(N) \over p(N)^n} 
\label{a2}
\end{equation}

Since for a given $N$ there is in general more than one 
solution to the extremum equation (\ref{ext}), here $p(N+1)$ and $p(N)$ 
will refer to solutions at consequtive orders which are both positions of minima or 
both positions of maxima. Now define 
\begin{equation}
p(N+1) = c(N) \ p(N) \,
\label{ac}
\end{equation}
where $c(N)$ is some function of $N$. 
Then from equation (\ref{a1}) and the corresponding one at order $N+1$,
one easily deduces
\begin{equation}
\lambda \ { \partial (S_{N+1}-S_{N}) \over \partial \lambda}|_{\lambda_0} =  \sum_{n=1}^{N} {n \over p(N)^n} \left({A_n(N+1) \over c(N)^n} - A_n(N) \right) + {(N+1) A_{N+1}(N+1) \over c(N)^{N+1} p(N)^{N+1}} \, .
\label{diff}
\end{equation}
Now for large $N$, $A_n(N) \sim A_n (N+1)$, and write this simply as $A_n$.
Also define, $\Delta S_N \equiv S_{N+1} - S_{N}$.
Assume now that $p(N)$ is large and $c(N)>1$. Also assume that all the coefficients 
$A_n$ are generically 
of the same order, or at least do not increase rapidly with $n$ (the factorial growth of $f_n$ 
has already been taken care of by the Borel transform). Then keeping terms needed to solve 
for $1/c(N)$ at leading order,
(\ref{diff}) simplifies to
\begin{equation}
\lambda \ { \partial \Delta S_{N} \over \partial \lambda}|_{\lambda_0} = {A_1  \over p(N)} \ \left({1 \over c(N)}- 1\right) - {2A_2 \over p(N)^2} \, .
\label{cn}  
\end{equation}
Similarly, from eq.(\ref{a2}) and the corresponding one at order $N+1$ one
deduces at large $N$ and to leading order in $1 / p(N)$,
\begin{equation}
\Delta S_N  =  {A_1 \over p(N)} \ \left({1 \over c(N)} -1\right) \, ,
\label{delta}
\end{equation}
where it is implicit in this equation that $\lambda= \lambda_0$. Now, if
$\lambda_0$ is varied, then the leading change in (\ref{delta}) comes   
from $A_1$. Thus using (\ref{delta}), Eq.(\ref{cn}) can be simplified to
\begin{equation}
{\alpha  \over p(N)} \ \left({1 \over c(N)}- 1\right) \approx {2A_2 \over p(N)^2} \, .
\label{cn1}  
\end{equation}
where $\alpha = (A_1 - \lambda {\partial A_{1} \over \partial \lambda})|_{\lambda_0}$. 
Thus,
\begin{equation}
{1 \over c(N)} \approx 1 +  {2 A_2 \over \alpha p(N)} \, .
\label{cn2}  
\end{equation}
This solution will be self-consistent with the initial 
assumption $c(N)>1$ only if $A_2 / \alpha $ is negative, 
which then requires that that $f_2$ and $f_1$ should be of opposite 
signs (since $A_1 $ is larger than ${\partial A_1 \over \partial \lambda})$. 
Note that for $N=2$ the solution $p(2)$ exists in the first place only 
if $f_2$ and $f_1$ are of opposite signs. Hence one concludes 
that $c(2)>1$ is generically expected. Comparing (\ref{cn2}) with the corresponding
equation at the next order $N+1$ one obtains at large $N$ the recursive relation
\begin{equation}
{1 \over c(N+1)} = 1- {1 \over c(N)} + {1 \over c(N)^2} \, ,
\label{recus}
\end{equation}
which can also be written as 
\begin{equation}
{1 \over c(N+1)} - {1 \over c(N)}  = \left(1- {1 \over c(N)} \right)^2 \,
\end{equation}
showing that 
\begin{equation}
c(N+1) < c(N) \, .
\label{dec}
\end{equation} 
Indeed the large $N$ solution of  (\ref{recus}) is 
\begin{equation}
{1 \over c(N)} \approx 1 + {1 \over N+K} \, .
\label{c1}
\end{equation}
with $K$ a constant.
Therefore $c( N \to \infty) \to 1^{+}$,
that is, though $p(N)$ will increase with $N$, it will approach a constant 
as $N \to \infty$. 
{ \it Now, since $p(N)$ increases with $N$, this means that the various approximations 
that led from (\ref{a01}) to (\ref{recus}-\ref{dec}) will become increasingly
accurate}. What is remarkable is that in the examples studied with $c(N)>1$, 
(\ref{dec}) is already 
satisfied at low $N$ and furthermore  the relation (\ref{recus}) too is a 
reasonable approximation.

The above relations (\ref{recus}-\ref{dec}) were obtained under the assumption 
$c(N) >1$. If $c(N)<1$ then there are apparently no simple relations. For the
auxiliary series
of the beta function in Sec.(3.5), $c(N)$ alternated between being 
slightly larger than one and much smaller than one.

Consider now the Eq.(\ref{delta}) which is valid at large 
$N$ and large $p(N)$,
\begin{equation}
\Delta S_N  =  {A_1 \over p(N)} \ \left({1 \over c(N)} -1\right) \, ,
\label{magic}
\end{equation}
Remarkably, this simple equation summarizes most of the observed trends.
When $c(N) >1$, (\ref{magic}) implies that
for $A_1 <0$, which corresponds to $f_1 <0$ and minima
solutions to (\ref{ext}),
 $\Delta S_N >0$,
which is indeed observed in the examples studied and leads to lower bounds. 
Similarly if $A_1 >0$, which corresponds to
$f_1 > 0$ and maxima solutions, $\Delta S_N <0$, implying upper bounds.
The only exception observed so far is the Euler-Heisenberg series 
in Sec.(3.4) which had $c(N)>1, \ A_1 <0$,
and yet gave upper bounds. Presumably for that example the 
approximation (\ref{magic}) is not
appropriate. 
Now suppose that $c(N)<1$, as happens roughly for the
auxiliary series in Sec.(3.5). Then Eq.(\ref{magic}) implies that 
minima, corresponding to $f_1 <0$ and hence $A_1 <0$,
give $\Delta S < 0$, which explains some of the oddities observed in \cite{unpub}.   

The equation (\ref{magic}) also explains the rapid convergence of the bounds. Indeed one sees
that convergence can be achieved simply by one of two ways. Firstly, if 
$c(N) \to 1$ as $N \to \infty$, that is $p(N) \to p_0$. This type
of behaviour, is manifested, for example, by the toy-model in Sec.(3.5) of Ref.\cite{unpub}. The second
way is for $c(N)>1$ for all $N$, 
which leads to $p(N)$
increasing with $N$. This is what has been generically observed. 
Actually, as shown above, the second behaviour also gives $c(N) \to 1^{+}$, 
and hence the convergence is doubly fast.
Explicitly, from Eqns.(\ref{cn2},\ref{c1},\ref{magic})
one deduces for the $c(N)>1$ case,
\begin{eqnarray}
\Delta S & \approx & {\alpha A_{1} \over 2 A_2} \left({1 \over c(N)}-1 \right)^2 \, \\
& = & {\alpha A_{1} \over 2 A_2} {1 \over (N+K)^2} \, .
\end{eqnarray}

In summary, for practical problems, 
the strongest statements can be made for the case when $c(N)>1$ is observed for the
given terms of a partial perturbative series, that is when $p(N)$ increases with
$N$. Then the various approximations leading to the above equations become
increasingly accurate at large $N$, allowing one to make assertions about all $N$. 
Firstly, one deduces $c(N \to \infty) \to 1^{+}$. Secondly, for $S_N$ which are 
global (and local) minima, if $S_N < S_{N+1}$ for the given terms of the series, 
the trend will continue for larger $N$, the convergence of the $S_N$ will be rapid,
and they will form lower bounds to the exact result. However if $S_N > S_{N+1}$ is observed 
for the global (and local) minima, (as in Sec.(3.4)), then though the trend will continue
and though the convergence of the $S_N$ will be rapid, it is not {\it a priori} obvious that
they will be upper bounds to the exact result, because then 
key pieces (\ref{star}-\ref{limit}) of the argument are missing. (The same 
loophole occurs for the approximants $\bar{S}_N$ formed from the $\bar{p}(N)$'s.)

If it is observed that $c(N)<1$ for the given terms of a series, then the various
equations above are not necessarily accurate at higher $N$. In that case one can only 
conjecture that the observed monotonic and rapid convergence will continue 
at higher orders.\\ \\

\noindent
{\large {\bf  A2}}

Consider now the slope of the approximants $S_N(\lambda)$.
From (\ref{a01}), as $\lambda \to 0$, $S_N(\lambda) \to f_0 + f_1 \lambda $, 
so that
\begin{equation}  
{\partial S_{N} \over \partial \lambda}|_{\lambda = 0} = f_1 \, .  
\label{slope0}
\end{equation}
Thus all the bounds approach the origin with the same value ($f_0$) and 
slope ($f_1$) independent of $p$ and $N$.This fact can be seen
in all the figures. Consider next Eq.(\ref{a1}) for large $p(N)$,
\begin{equation}
{ \partial S_{N} \over \partial \lambda}|_{\lambda_0} \sim  {1 \over \lambda_0} {A_1 \over p(N)} \, .
\label{slope1}
\end{equation}
Since the sign of $A_1$ is the same as the sign of $f_1$, this shows that
the curves $S_N(\lambda)$ are not expected to change direction as $\lambda$
varies. This is indeed observed, and explains why
the bounds $S_N(\lambda)$ to $S(\lambda)$
are also good estimates of $S(\lambda)$
itself only when the latter has a slowly
varying first derivative. 

As was observed in Sec.(3.5) (see also \cite{unpub}), the approximants $\bar{S}(\lambda)$ formed from
the local extrema $\bar{p}(N)$ had a more varying slope. 
In order to understand this, set for simplicity $f_0=0$ 
so that $\bar{S}_{N}(0)=0$ and let us demand that
\begin{equation}
\bar{S}_N(\lambda_0) =0 \, ,
\label{demand}
\end{equation}
so that $\bar{S}_N(\lambda)$ curves back to its value at the origin, 
and so can better approximate functions like $\sin(\lambda \pi).$ 
The condition (\ref{demand}) then leads from (\ref{a02}) to 
\begin{equation}
0 = \sum_{n=1}^{N} {A_n \over \bar{p}(N)^n} \, .
\label{demand2}
\end{equation}
Note that since the $\bar{p}(N)$ also have to satisfy the
extremum condition (\ref{ext}), eqns.(\ref{ext},\ref{demand2}) are actually
two coupled equations for $\lambda_0$ and $\bar{p}(N)$ which we would like to analyse
for consistency. Firstly, (\ref{demand2}) can 
be used to eliminate the leading term in the 'barred' version of (\ref{a1}),
so that now
\begin{equation}
\lambda \ { \partial \bar{S}_{N} \over \partial \lambda}|_{\lambda_0} = \sum_{n=2}^{N} {(n-1) A_n \over p(N)^n} \, 
\label{a11}
\end{equation}
which at large $\bar{p}(N)$ gives the slope  
\begin{equation}
{ \partial \bar{S}_{N} \over \partial \lambda}|_{\lambda_0} \sim  {1 \over \lambda_0} {A_2 \over p(N)^2} \, .
\label{slope2}
\end{equation}
If $f_2$ is opposite in sign to $f_1$ then  by comparing ($\ref{slope2})$ 
with (\ref{slope0}) one sees that indeed the approximants $\bar{S}$ can change direction
Of course this just shows that the approximate large $\bar{p}(N)$ 
analysis above is self-consistent.   
Now, in Sec.(2), for $f_1 <0, \ f_2 >0$, 
the $\bar{p}(N)$ have been defined as positions of local maxima when
the $p(N)$ are positions of local minima. We can compare the 
relative magnitudes of the two values as follows. For 
large $\bar{p}(N)$, (\ref{demand2}) has the approximate solution 
\begin{equation}
\bar{p}(N) \sim { -A_2 \over A_1} \, .
\end{equation}
By contrast the $p(N)$ are approximate solutions of (\ref{a1}) with the left-hand-side
deleted, which is equivalent to ignoring the mild $p$ dependence of
the $A_n$'s,
\begin{equation}
p(N) \sim {-2 A_2 \over A_1} \, .
\end{equation}
Thus the values of $\bar{p}(N)$ are expected to be smaller than those
of $p(N)$, and the interested reader may verify from the given examples
that this is indeed the case. 
Reversing the logic of the argument above, one concludes as follows.
For that $N$ when $p(N)$ is the position of a local minimum,
one expects a local maximum at $\bar{p}(N)$. While the $S_N$ curve is
expected to be monotonic in $\lambda$, that of $\bar{S}(\lambda)$ 
will not be monotonic if the value of $\bar{p}(N)$ is smaller than that 
of $p(N)$.

A similar discussion can be carried out for the curves formed 
from an auxiliary series $S_{N}^{'}$.
Again set for simplicity $f_0=0$, and define  
\begin{equation}
\hat{S}^{'}_N \equiv {\hat{S}^{'}_N \over \lambda} \, .
\label{relation}
\end{equation}
The extremization (\ref{ext}) in $p$ is done with respect to
the resummed auxiliary series $S^{'}_N$, and one deduces for large
$p^{'}(N)$, from an equation analogous to (\ref{a1}), that
\begin{equation}
\mbox{sign}\left( \lambda {\partial S^{'}_N \over \partial \lambda}\right)_{\lambda_0} = \mbox{sign} (f_2) \, .
\label{sign1}
\end{equation}
Now demanding
\begin{equation}
S_N(\lambda_0) =0 \, ,
\label{demand3}
\end{equation}
for the 
reconstructed resummed series $S_N \equiv \lambda S^{'}_N$ gives,
\begin{equation}
 {\partial S_N \over \partial \lambda}|_{\lambda_0} = \lambda_0 \ \left({\partial S^{'}_N \over \partial \lambda}\right)_{\lambda_0}  \, ,
\label{sign2}
\end{equation}
which when combined with (\ref{sign1}) shows that for $f_1$ and $f_2$ of
opposite signs, the slope of the reconstructed 
$S_N$ at $\lambda= \lambda_0$ is opposite in sign to
its slope at the origin which is given by (\ref{slope0}).   
  
Thus if $f_1$ and $f_2$ are of opposite signs, then the auxiliary series
may be expected to give a {\it reconstructed} $S_N$ with a slope that 
has variation in sign, compared with the slope of the approximant 
$S_N$ which is obtained by direct means, if the two conditions 
(\ref{ext}) and (\ref{demand3}) for $S^{'}$ have a consistent solution
$p'(N)$ for some $\lambda_0$.

\newpage

{\bf Figure Captions}

Figure (1a): Plots of the divergent series $\hat{S}_N(\lambda)$ for the model
in Sec.(3.1),
together with the exact result $S(\lambda)$. Starting from the lowest
curve and moving upwards, one has $\hat{S}_5, \ \hat{S}_3, \ S, \ \hat{S}_2, \ \hat{S}_4$.
\\

Figure (1b):
Plots of the resummed series $S_N(\lambda)$ for the model in Sec.(3.1), together
with the exact result $S(\lambda)$. Starting from the lowest curve and moving upwards,
one has $S_2, \ S_3, \ S_4, \ S$. 
\\

Figure (1c):
Plots of the resummed series $S_{(0)N} (\lambda)$ for the model in Sec.(3.1), together
with the exact result $S(\lambda)$. Starting from the lowest curve and moving upwards,
one has $S_4, \ S, \ S_5, \ S_3$. Compared to Fig.(1b), the convergence is faster but not
monotonic.
\\

Figure (2a): Plots of the series $\hat{S}_N(\lambda)$ for the model
in Sec.(3.2),
together with the exact result $S(\lambda)$. Starting from the lowest
curve and moving upwards, one has $\hat{S}_3, \ S, \ \hat{S}_4, \ \hat{S}_2$.  
\\

Figure (2b):
Plots of the resummed series $S_N(\lambda)$ for the model in Sec.(3.2), together
with the exact result $S(\lambda)$. Starting from the lowest curve and moving upwards,
one has $S_2, \ S_3, \ S_4, \ S$. 
\\

Figure (3a):
Plots of the resummed perturbation series $S_N(\lambda)$ for the model in Sec.(3.3), together
with the full perturbative result $S_{pert}(\lambda)$.
Starting from the lowest curve and moving upwards,
one has $S_2, \ S_3, \ S_4, \ S_5, \ S_{pert}$. 
\\

Figure (3b): Plots of $S_{exact}(\lambda), \ S_5(\lambda)$ and $S_{5}(\lambda) -S_{np}(\lambda)$
as defined in Sec.(3.3). At $\lambda=10$, the lowest curve is of $S_5 -S_{np}$, and the 
highest one is $S_5$.  
\\

Figure (4):
Plots of the resummed Euler-Heisenberg series $S_N(\lambda)$, together
with Schwinger's exact result $S(\lambda)$. Starting from the lowest curve 
and moving upwards, one has $S, \ S_6, \ S_4, \ S_2$. The approximants approach the 
exact result from above.
\\

Figure (5a): Plots of the resummed beta function ${S}_N(\lambda)$ of
Sec.(3.5). Starting from the lowest
curve and moving upwards, one has $S_7, \ S_6, \ S_5, \ S_4, \ S_3, S_2$. 
\\

Figure (5b):
Plots of the resummed series $\bar{S}_N(\lambda)$ for beta function in Sec.(3.5).
Starting from the lowest curve and moving upwards,
one has $\bar{S}_3, \ \bar{S}_5, \ \bar{S}_7$. The approximants appear to form upper 
bounds.
\\

Figure (5c): Plots of the resummed beta function ${S}_N(\lambda)$ of
Sec.(3.5), obtained through the auxiliary series. Starting from the lowest
curve and moving upwards, one has $S_2, \ S_3, \ S_4, \ S_5, \ S_6$. 
The approximants appear to form upper bounds. 
The curves for $N=2$ and $N=3$ are indistinguishable, and similarly, those for
$N=4$ and $N=5$ are very close. 
\\

Figure (5d):
Magnification of the $N=6$ curve of Fig.(3.5) near its nontrivial zero.
\\

\newpage
\input{epsf.sty}

\epsfbox{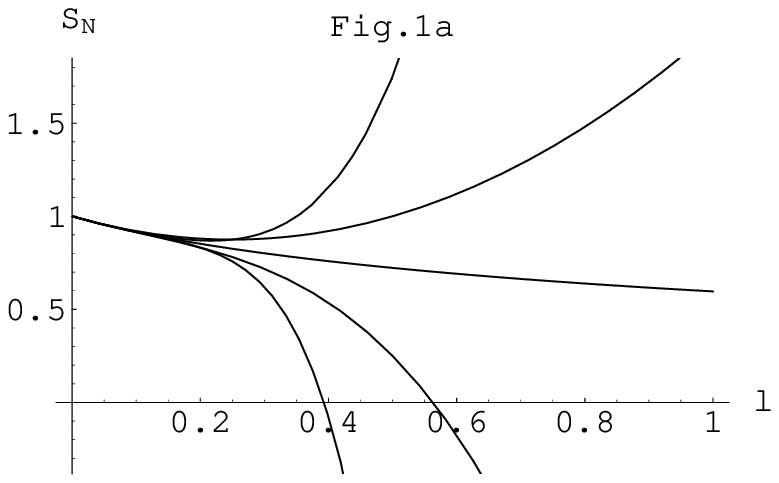}
\vspace{0.5in}

\epsfbox{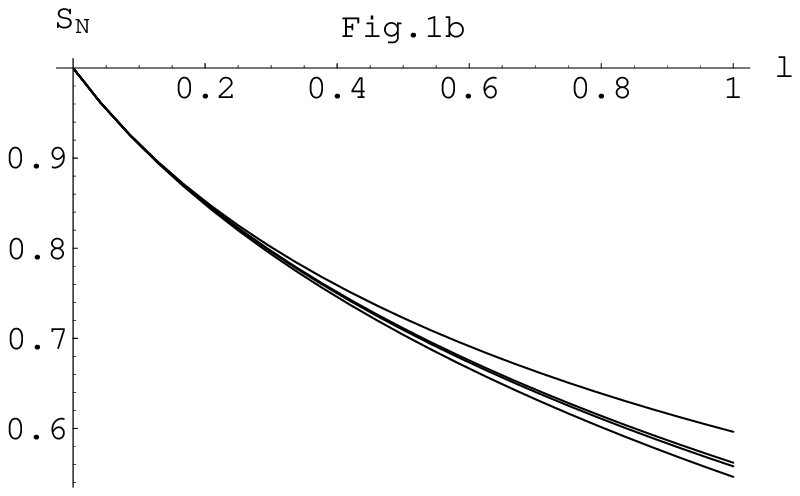}
\vspace{0.5in}

\epsfbox{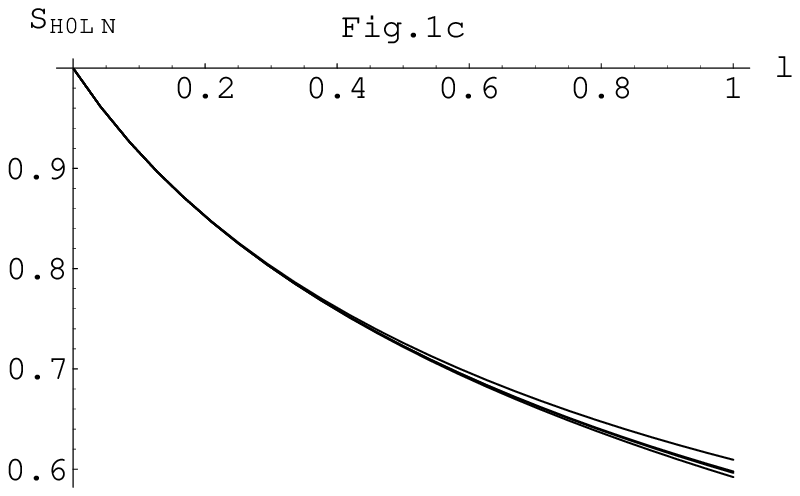}
\vspace{0.5in}

\epsfbox{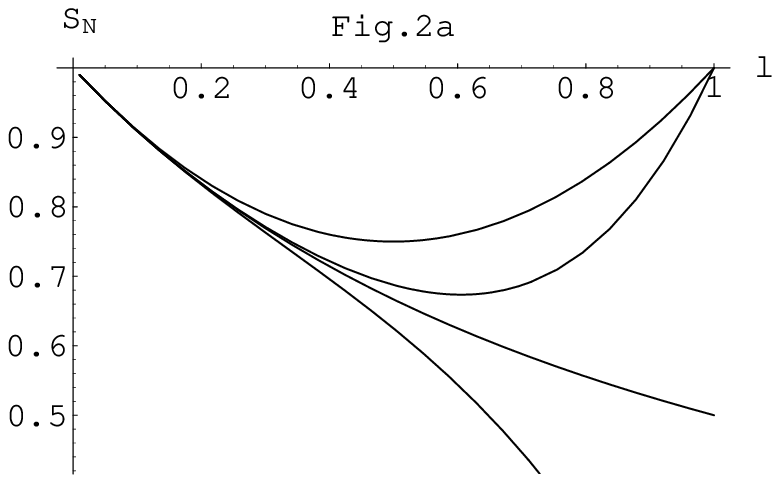}
\vspace{0.5in}

\epsfbox{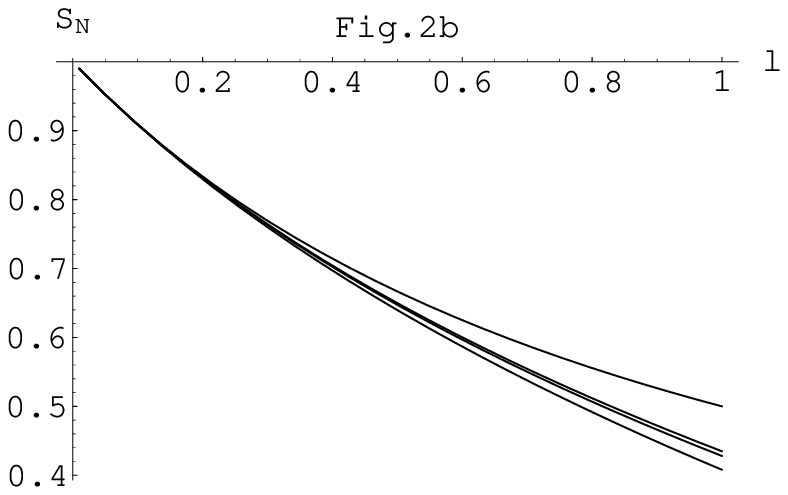}
\vspace{0.5in}

\epsfbox{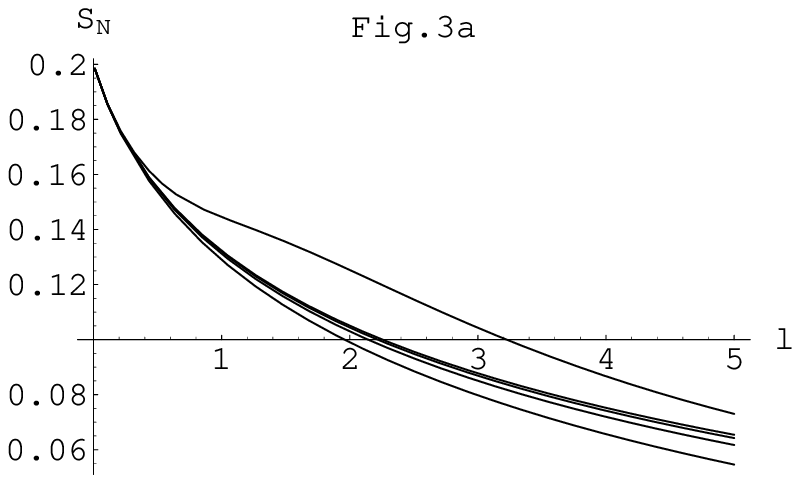}
\vspace{0.5in}

\epsfbox{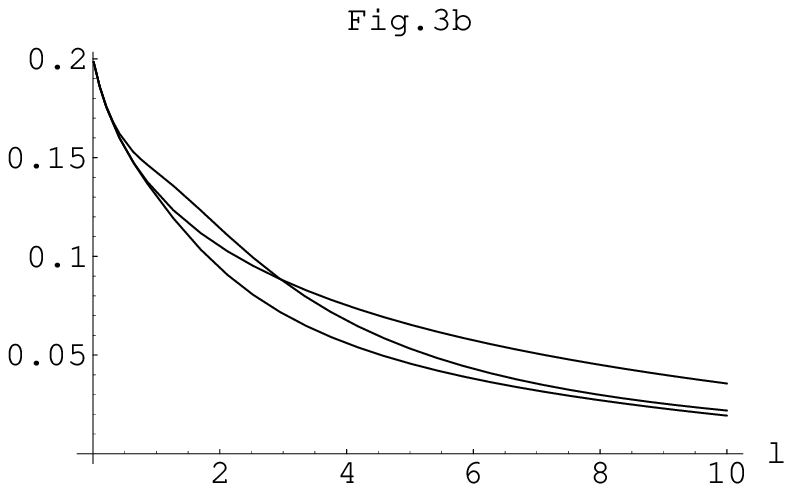}
\vspace{0.5in}

\epsfbox{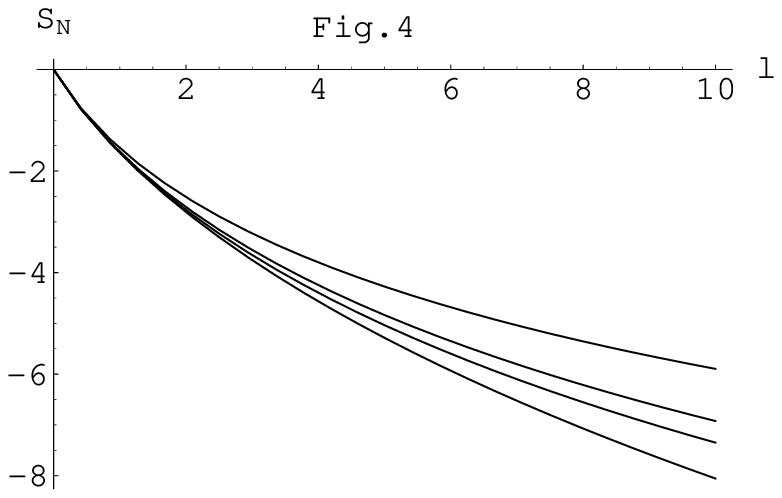}
\vspace{0.5in}

\epsfbox{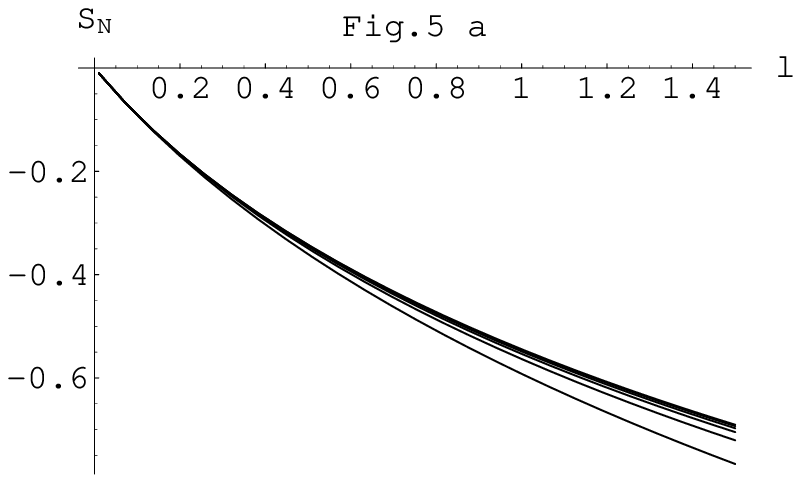}
\vspace{0.5in}

\epsfbox{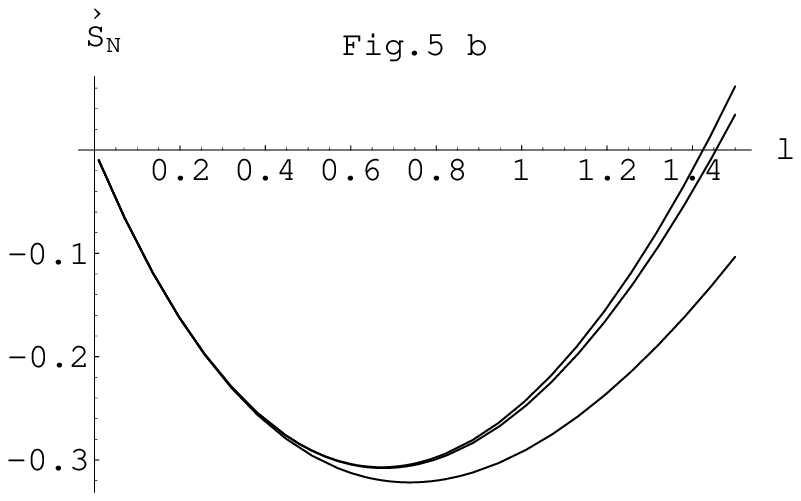}
\vspace{0.5in}

\epsfbox{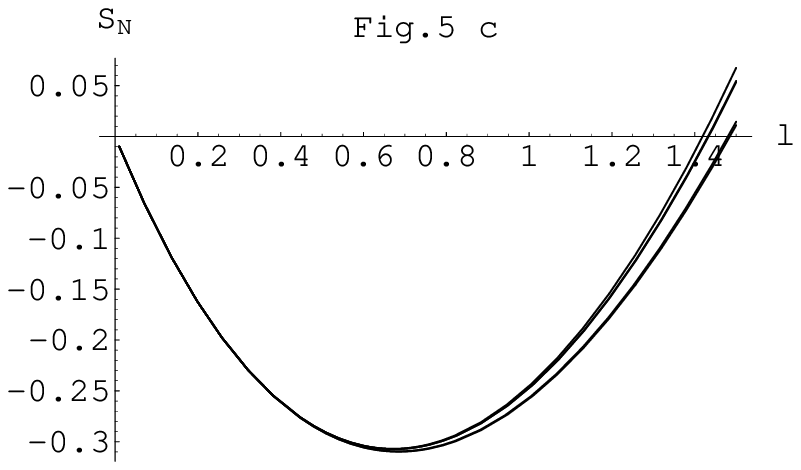}
\vspace{0.5in}

\epsfbox{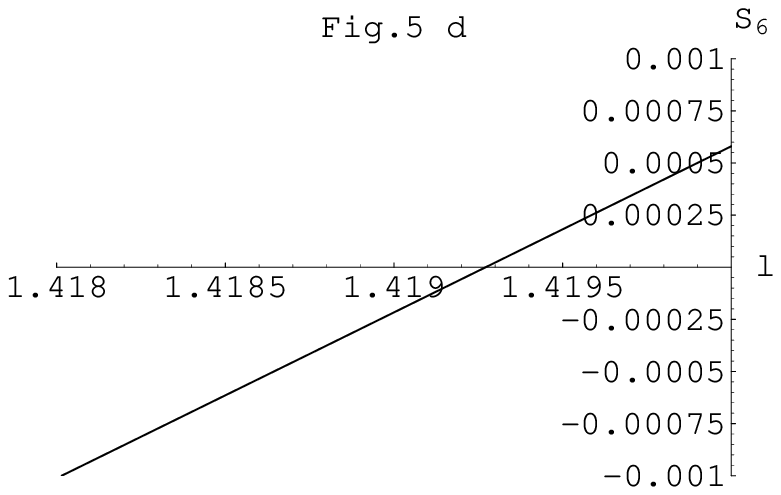}

\end{document}